\documentclass[12pt]{article}
\usepackage{epsfig}
\usepackage{amssymb}
\def\ie{\textsl{i.e.}}

\newcommand{\comment}[1]  {  }
\def\BE{\begin{equation}}
\def\EE{\end{equation}}
\def\BEA{\begin{eqnarray}}
\def\EEA{\end{eqnarray}}
\newcommand{\pd}[2]{\frac{\partial #1}{\partial #2}}
\newtheorem{thm}{Theorem}
\newcommand\vb{{\bf b}}

\newcommand\mD{{\bf D}}

\newcommand\mR{{\bf R}}

\begin{document}
\baselineskip 12pt

\begin{center}
{\huge\bf Optimum Asymptotic Multiuser Efficiency of Pseudo-Orthogonal Randomly Spread CDMA} \\
\vspace{5mm}
\begin{tabular}{cc}  Ori Shental & Ido Kanter  \\
Tel-Aviv University & Bar-Ilan University \\
Dept. of Electrical Engineering-Systems & Dept. of Physics \\
Tel-Aviv 69978 & Ramat-Gan 52900\\
Israel & Israel \\
\verb+shentalo@eng.tau.ac.il+ & \verb+kanter@mail.biu.ac.il+
\end{tabular}
\end{center}

\begin{abstract}
A $K$-user pseudo-orthogonal (PO) randomly spread CDMA system,
equivalent to transmission over a subset of $K'\leq K$ single-user
Gaussian channels, is introduced. The high signal-to-noise ratio
performance of the PO-CDMA is analyzed by rigorously deriving its
asymptotic multiuser efficiency (AME) in the large system limit.
Interestingly, the $K'$-optimized PO-CDMA transceiver scheme
yields an AME which is practically equal to 1 for system loads
smaller than $0.1$ and lower bounded by 1/4 for increasing loads.
As opposed to the vanishing efficiency of linear multiuser
detectors, the derived efficiency is comparable to the ultimate
CDMA efficiency achieved for the intractable optimal multiuser
detector.
\end{abstract}

\section{Introduction}
Non-orthogonality of realistic spreading sequences, serving as
signature codes in direct-sequence code division multiple-access
(DS-CDMA), has an inherent detrimental effect in cellular
communications. Nevertheless, a de-facto orthogonality between
transmissions can be achieved via a proper joint binary signaling,
for which multiple-access interference (MAI) agrees with
information polarity.

By borrowing analysis tools from statistical mechanics, counting
the number of metastable states of the Hopfield model of neural
networks~\cite{BibDB:Gardner}, we rigorously derive the asymptotic
multiuser efficiency (AME,~\cite{BibDB:BookVerdu}) of this
pseudo-orthogonal modulation scheme for the common random
spreading CDMA channel and draw its optimum value.

\section{Pseudo-Orthogonal Random CDMA}
Consider a perfectly power-controlled, synchronous, $K$-users,
$N$-chips random binary spreading, binary signaling, Gaussian CDMA
channel. A random CDMA channel is termed 'Pseudo-Orthogonal' (PO)
if the transmitted signaling vectors
$\{-1,1\}^{K}\ni\vb\triangleq\{b_{1},\ldots,b_{K}\}$ are chosen
such that \BE\label{eq_PO} \mR\vb\equiv\mD\vb, \EE or in scalar
form \BE\label{eq_PO2}
\sum_{i=1}^{K}\rho_{ki}b_{i}\equiv\lambda_{k}b_{k}\quad\forall
k=1\ldots K, \EE where the $K\times K$ symmetric matrix $\mR$,
with entries $\rho_{ki}$, is the spreading sequences' normalized
crosscorrelation matrix, and the $K\times K$ matrix \BE
\mD\triangleq\textrm{diag}\{\lambda_{1}>0,\ldots,\lambda_{K'}>0,\lambda_{K'+1}>-\infty,\ldots,\lambda_{K}>-\infty\}
\EE is a diagonal matrix, in which the first $K'\leq K$ diagonal
entries must be positive and the rest can get any value.

To put differently, the allowed signaling vectors $\vb$ are those
for which the \emph{MAI does not flip the information bits} of a
given fraction $\gamma\triangleq K'/K$ of the users. Evidently,
this pseudo-orthogonality is gained at the cost of reducing the
signaling entropy per user, $H$, to be less than unity (as opposed
to multiuser precoding~\cite{BibDB:Precoding} in which there is no
restriction on the input signaling.)

On the other hand, this scheme reverts the multiuser channel to an
equivalent set of $\gamma H$ single-user additive white Gaussian
noise (AWGN) channels, with much simpler detectors w.r.t. the
intractable optimal multiuser detector (MUD). This joint signaling
scheme is particularly attractive for the CDMA downlink (but may
also be utilized for the uplink in case of cooperation between
users.) In order to evaluate the high signal-to-noise ratio (SNR)
bit-error performance of the PO-CDMA scheme, its AME is analyzed.

\section{Asymptotic Multiuser Efficiency}
\begin{thm}
In the large-system limit analysis,\,\ie, $K,N\rightarrow\infty$,
yet the system load factor $\beta\triangleq K/N$ is kept constant,
the PO-CDMA channel's AME, which characterizes the performance
loss (in effective SNR) as the background noise
vanishes~\cite{BibDB:BookVerdu}, is proved (for a given active
users fraction $\gamma$) to get the form \BEA
    \eta(\beta,\gamma)&=&\frac{\gamma\log_{2}e}{\beta}\big(b^{\ast}-\frac{1}{2}+\frac{(1-b^{\ast})^{2}}{2a^{\ast}}+\frac{1}{2}\log{a^{\ast}}\big)\nonumber\\&+&\gamma^{2}\log_{2}\big(2Q(t^{\ast})\big)+\gamma(1-\gamma)\nonumber,
\EEA where $Q(\cdot)$ is the complementary cumulative distribution
function of a standard Gaussian random
variable\footnote{$Q(x)\triangleq1/\sqrt{2\pi}\int_{x}^{\infty}dy\exp{(-y^{2}/2)}$},
and the fixed-points $a^{\ast}$, $b^{\ast}$ and $t^{\ast}$ are
devised by solving numerically the following set of equations \BEA
    \beta^{-1}\big(\frac{(1-b)^{2}}{a}-1\big)+\gamma t\frac{Q'(t)}{Q(t)}=0,\nonumber\\
    \beta^{-1}\big(1-\frac{1-b}{a}\big)+\frac{\gamma}{\sqrt{a\beta}}\frac{Q'(t)}{Q(t)}=0\nonumber,
\EEA with an auxiliary variable $t\triangleq(b-1)/\sqrt{a\beta}$.
\end{thm}

\section{Proof}
If a particular multiple-access system achieves bit-error-rate
(BER) per user $P(\sigma)$ in the presence of MAI and AWGN with
power spectral level equal to $\sigma^{2}$, then the AME w.r.t. a
single-user system (or a fully-orthogonal CDMA) is defined
by~\cite{BibDB:BookVerdu}\BE\label{eq_etadef}
    \eta\triangleq\lim_{\sigma\rightarrow0}e(\sigma).
\EE The term $e(\sigma)$ is the energy per user required to
achieve BER equal to $P(\sigma)$ in a single-user Gaussian channel
with the same background noise level.

According to definition~(\ref{eq_etadef}), the AME for the PO-CDMA
case~(\ref{eq_PO}), $\eta(\beta,\gamma)$, is nothing but $\gamma
H(\beta,\gamma)$, the fraction of 'orthogonal' users multiplied by
their signaling entropy (as one is interested in an equivalent set
of single-user channels with unit entropy.) Thus we have to
compute the non-trivial entropy $H(\beta,\gamma)$ under the
PO-CDMA transmission constraints~(\ref{eq_PO2}).

A $K$-length binary transmission codeword
$\vb^{c}\triangleq\{b_{1}^{c},\ldots,b_{K}^{c}\}$, composed of all
$K$ users' bits at a given channel use, for which the PO
constraints~(\ref{eq_PO2}) hold, satisfies the condition
\BE\label{eq_stable}
    \int_{0}^{\infty}\prod_{k=1}^{K'}d\lambda_{k}\delta(\sum_{i}\rho_{ki}b_{i}-\lambda_{k}b_{k})=\int_{-1}^{\infty}\prod_{k=1}^{K'}d\lambda_{k}\delta(\sum_{i\neq
    k}\rho_{ki}b_{i}-\lambda_{k}b_{k})=1,
\EE where the function $\delta(\cdot)$ is the Dirac delta
function. Condition~(\ref{eq_stable}) can be reformulated as \BEA
    &&\alpha\int_{-1}^{\infty}\prod_{k=1}^{K'}d\lambda_{k}\delta(\sum_{i\neq
    k}\alpha\rho_{ki}b_{i}-\alpha\lambda_{k}b_{k})=\nonumber\\
    &&\int_{-\alpha}^{\infty}\prod_{k=1}^{K'}d\lambda_{k}\delta(\sum_{i\neq
    k}\alpha\rho_{ki}b_{i}-\lambda_{k}b_{k})=1,
\EEA where $\alpha\triangleq1/\beta$.

Let the random variable $\mathbb{N}(\beta,\gamma,\mR)$ denote the
number of PO-CDMA allowable codewords, \ie \BE\label{eq_number0}
    \mathbb{N}(\beta,\gamma,\mR)\triangleq\\\int_{-\alpha}^{\infty}\prod_{k=1}^{K'}d\lambda_{k}\delta(\sum_{i\neq k}\alpha\rho_{ki}b_{i}-\lambda_{k}b_{k}).
\EE Assuming equal user information rates, the corresponding
asymptotic signaling entropy is defined~\cite{BibDB:BookCover}, in
bit information units, as \BE\label{eq_capacity}
    H(\beta,\gamma)\triangleq\lim_{K\rightarrow\infty}\frac{\log_{2}{\mathbb{N}(\beta,\gamma,\mR)}}{K}.
\EE Assuming self-averaging property~\cite{BibDB:BookEllis}, in
the large-system limit $K\rightarrow\infty$ the number of
successful codewords $\mathbb{N}(\beta,\gamma,\mR)$ is equal to
its expectation w.r.t. the distribution of $\mR$, \ie
\BEA\label{eq_number1}
    &&\lim_{K\rightarrow\infty}\mathbb{N}(\beta,\gamma,\mR)=\mathcal{N}(\beta,\gamma)\nonumber\\&=&
    \lim_{K\rightarrow\infty}\int_{-\alpha}^{\infty}\prod_{k=1}^{K'}d\lambda_{k}
    \sum_{\vb}\Bigg<\prod_{k=1}^{K'}\delta\Big(\sum_{i\neq k}\alpha\rho_{ki}b_{i}-\lambda_{k}b_{k}\Big)\Bigg>_{\mR},
\EEA where $\mathcal{N}(\beta,\gamma)$ and $<\cdot>_{\mR}$ denote
the average and averaging operation w.r.t. $\mR$, respectively,
and $\sum_{\vb}$ corresponds to a sum over all the $2^K$ possible
values of $\vb$.

Representing the delta function by the inverse Fourier transform
of an exponent, expression~(\ref{eq_number1}) can be rewritten as
\BEA\label{eq_number2}
    \mathcal{N}(\beta,\gamma)&=&\lim_{K\rightarrow\infty}\int_{-\alpha}^{\infty}\prod_{k=1}^{K'}d\lambda_{k}
    \frac{1}{(2\pi)^{K'}}\int_{-\infty}^{\infty}\prod_{k=1}^{K'}d\omega_{k}\nonumber\\
    &\times&\sum_{\vb}\exp{\Big(j\sum_{k=1}^{K'}{\omega_{k}\lambda_{k}b_{k}}\Big)}
    \Bigg<\exp{\Big(-j\sum_{i\neq
    k}\alpha\rho_{ki}b_{i}\omega_{k}\Big)}\Bigg>_{\mR},
\EEA where $j\triangleq\sqrt{-1}$ and $\omega$ is the angular
frequency of the Fourier transform. Substituting $b_{k}\omega_{k}$
for $\omega_{k}$, we find \BEA\label{eq_number3}
    \mathcal{N}(\beta,\gamma)&=&\lim_{K\rightarrow\infty}\int_{-\alpha}^{\infty}\prod_{k}d\lambda_{k}
    \frac{1}{(2\pi)^{K}}\int_{-\infty}^{\infty}\prod_{k}d\omega_{k}\nonumber\\
    &\times&\sum_{\vb}\exp{\Big(j\sum_{k}{\omega_{k}\lambda_{k}}\Big)}
    \cdot\mathbb{E},
\EEA where \BEA\label{eq_expectation}
    \mathbb{E}&\triangleq&\Bigg<\exp{\Big(-j\sum_{i\neq k}\alpha\rho_{ki}b_{i}b_{k}\omega_{k}\Big)}\Bigg>_{\mR}\nonumber\\
    &=&\Bigg<\exp{\Big(-j\sum_{i\neq k}\frac{1}{K}\sum_{\mu=1}^{N}s_{k}^{\mu}s_{i}^{\mu}b_{i}b_{k}\omega_{k}\Big)}\Bigg>_{\mR}
. \EEA In the last equality the cross-correlations
$\rho_{ki}\triangleq1/N\sum_{\mu=1}^{N}s_{k}^{\mu}s_{i}^{\mu}$ are
expressed explicitly as a function of the binary chips
$s_{k}^{\mu}=\pm1$. The expectation $\mathbb{E}$ can be also
written as \BEA\label{eq_expectation11}
    \mathbb{E}&=&\exp{(j\alpha\sum_{k}\omega_{k})}\Bigg<\exp{\Big(-\frac{j}{K}\sum_{\mu}(\sum_{k}s_{k}^{\mu}b_{k}\omega_{k})(\sum_{k}s_{k}^{\mu}b_{k})\Big)}\Bigg>_{\mR}.
\EEA

Using the transformation~\cite{BibDB:BruceEtAl} \BEA
    \exp{(-jA_{\mu}B_{\mu}/K)}&=&\int_{-\infty}^{\infty}\frac{da_{\mu}}{(2\pi/K)^{1/2}}
    \int_{-\infty}^{\infty}\frac{db_{\mu}}{(2\pi/K)^{1/2}}\\&\times&
    \exp{\Big(j\frac{K}{2}(a_{\mu}^{2}-b_{\mu}^{2})-\frac{j}{\sqrt{2}}A_{\mu}(a_{\mu}+b_{\mu})-\frac{j}{\sqrt{2}}B_{\mu}(a_{\mu}-b_{\mu})\Big)}\nonumber,
\EEA expression~(\ref{eq_expectation11}) becomes (here, and
hereafter, logarithms are taken to base
$e$)\BEA\label{eq_expectation2}
    \mathbb{E}&=&\exp{(j\alpha\sum_{k}\omega_{k})}\int_{-\infty}^{\infty}\prod_{\mu}\frac{da_{\mu}}{(2\pi/K)^{1/2}}
    \int_{-\infty}^{\infty}\prod_{\mu}\frac{db_{\mu}}{(2\pi/K)^{1/2}}\nonumber\\&\times&\exp{\Big(j\frac{K}{2}\sum_{\mu}(a_{\mu}^{2}-b_{\mu}^{2})
    +\sum_{k,\mu}\log\big(\cos(c_{k,\mu})\big)\Big)},
\EEA where \BE
    c_{k,\mu}\triangleq\frac{1}{\sqrt{2}}\big(\omega_{k}(a_{\mu}+b_{\mu})+(a_{\mu}-b_{\mu})\big).
\EE

Since $\sum_{k}s_{k}^{\mu}b_{k}$ in~(\ref{eq_expectation11}) is
$\mathcal{O}(\sqrt{K})$ for a vast majority of codewords, for the
expectation $\mathbb{E}$ to be finite, $a_{\mu}$ and $b_{\mu}$
must be $\mathcal{O}(1/\sqrt{K})$. Hence, expanding the
$\log\big(\cos(\cdot)\big)$ term in
exponent~(\ref{eq_expectation2}) and neglecting terms of order
$1/K$ and higher, we get \BEA\label{eq_expectation3}
    \mathbb{E}&=&\exp{(j\alpha\sum_{k}\omega_{k})}\int_{-\infty}^{\infty}\prod_{\mu}\frac{da_{\mu}}{(2\pi/K)^{1/2}}
    \int_{-\infty}^{\infty}\prod_{\mu}\frac{db_{\mu}}{(2\pi/K)^{1/2}}\nonumber\\&\times&\exp{\Big(j\frac{K}{2}\sum_{\mu}(a_{\mu}^{2}-b_{\mu}^{2})
    -\frac{1}{4}\sum_{k,\mu}\hat{c}_{k,\mu}\Big)}, \EEA
where \BE
    \hat{c}_{k,\mu}\triangleq\big(\omega_{k}^{2}(a_{\mu}+b_{\mu})^{2}+2\omega_{k}(a_{\mu}^{2}-b_{\mu}^{2})+(a_{\mu}-b_{\mu})^{2}\big).
\EE

Now, the solution of the $K$-dimensional
integral~(\ref{eq_expectation3}) of the expectation $\mathbb{E}$
is performed using the following mathematical recipe: New
variables are introduced\BEA
    a&\triangleq&\frac{1}{2\alpha}\sum_{\mu}(a_{\mu}+b_{\mu})^{2}\label{eq_a},\\
    b&\triangleq&\frac{j}{2\alpha}\sum_{\mu}(a_{\mu}^{2}-b_{\mu}^{2})+1\label{eq_b}.
\EEA Equations~(\ref{eq_a}) and~(\ref{eq_b}) can be reformulated
via the integral representation of a delta function using the
corresponding angular frequencies $A$ and $B$, respectively, \BEA
    \int_{-\infty}^{\infty}\frac{da\,dA}{2\pi/K\alpha}\exp{\big(jKA(\alpha
    a-\sum_{\mu}\frac{(a_{\mu}+b_{\mu})^{2}}{2})\big)}&=&1,\label{eq_aF}\\
    \int_{-\infty}^{\infty}\frac{db\,dB}{2\pi/K\alpha}\exp{\big(jKB(\alpha
    b-j\sum_{\mu}\frac{(a_{\mu}^{2}-b_{\mu}^{2})}{2}-\alpha)\big)}&=&1.\label{eq_bF} \EEA

Substituting these (unity) integrals into the expectation
expression~(\ref{eq_expectation3}) and rewriting it using $a$ and
$b$, the integrations over $a_{\mu}$ and $b_{\mu}$ are decoupled
and can be performed easily. Next, for the asymptotics
$K\rightarrow\infty$, the integration over the frequencies $A$ and
$B$ can be performed algebraically by the saddle-point
method~\cite{BibDB:BookEllis}.

According to this method, the main contribution to the integral
comes from values of $A$ and $B$ in the vicinity of the maximum of
the exponent's argument. Finally, the $\mathbb{E}$ term boils down
to \BEA\label{eq_expectation_final}
    \mathbb{E}&=&\int_{-\infty}^{\infty}\frac{da\
    db}{4\pi/K\alpha}\exp{\big(-\frac{1}{2}\alpha a\sum_{k}\omega_{k}^{2}+j\alpha b\sum_{k}\omega_{k}\big)}\nonumber\\
    &\times&\exp{\big(K\alpha(b-\frac{1}{2}+\frac{(1-b)^{2}}{2a}+\frac{1}{2}\log{a})\big)}.
\EEA

Substituting the expectation term~(\ref{eq_expectation_final})
back in~(\ref{eq_number3}), the integrand in the latter becomes
independent of $\vb$, and therefore the $\sum_{\vb}$ can be
substituted by multiplying with the scalar $2^K$. Hence, \BEA
    \mathcal{N}(\beta,\gamma)&=&\lim_{K\rightarrow\infty}\int_{-\alpha}^{\infty}\prod_{k}d\lambda_{k}\frac{2^{K-K'}}{\pi^{K'}}
    \int_{-\infty}^{\infty}\prod_{k}d\omega_{k}\exp{\Big(j\sum_{k}{\omega_{k}\lambda_{k}}\Big)}
    \nonumber\\&\times&\int_{-\infty}^{\infty}\frac{da\
    db}{4\pi/K\alpha}\exp{\Big(K\alpha\big(b-\frac{1}{2}+\frac{(1-b)^{2}}{2a}+\frac{1}{2}\log{a}\big)\Big)}
    \nonumber\\&\times&\exp{\Big(-\frac{1}{2}\alpha a\sum_{k}\omega_{k}^{2}+j\alpha b\sum_{k}\omega_{k}\Big)},
\EEA where the resulting $\omega$ dependent integrand is a
Gaussian function. Thus performing Gaussian integration and
exploiting the symmetry in the $K$-dimensional space, we get
\BEA\label{eq_number4}
    \mathcal{N}(\beta,\gamma)&=&\lim_{K\rightarrow\infty}\frac{2^{K-K'}}{\pi^{K'}}
    \int_{-\infty}^{\infty}\frac{da\
    db}{4\pi/K\alpha}\exp{\Big(K\alpha\big(b-\frac{1}{2}+\frac{(1-b)^{2}}{2a}+\frac{1}{2}\log{a}\big)\Big)}
    \nonumber\\&\times&\exp{\bigg(K'\log{\Big(\sqrt{\frac{2\pi}{\alpha a}}\int_{-\alpha}^{\infty}d\lambda\exp{\big(-\frac{(\alpha b+\lambda)^{2}}{2\alpha
    a}\big)}\Big)}\bigg)}.\EEA
Using the rescaling $(\alpha b+\lambda)/\sqrt{\alpha a}\rightarrow
\lambda$, the integral~(\ref{eq_number4}) becomes
\BE\label{eq_integral}
    \mathcal{N}(\beta,\gamma)=\lim_{K\rightarrow\infty}\int_{-\infty}^{\infty}\frac{da\
    db}{4\pi/K\alpha}\exp{\big(Kg(a,b,\gamma,\beta)\big)},
\EE where the function $g(a,b,\gamma,\beta)$ is defined by \BEA
    g(a,b,\gamma,\beta)\triangleq\frac{1}{\beta}\big(b-\frac{1}{2}+\frac{(1-b)^{2}}{2a}+\frac{1}{2}\log{a}\big)+\gamma\log\big(2Q(t)\big)+(1-\gamma)\log(2),
\EEA with an auxiliary variable \BE\label{eq_auxt}
    t\triangleq\frac{\sqrt{\alpha}(b-1)}{\sqrt{a}}.
\EE

Again, for $K\rightarrow\infty$, the double integral
in~(\ref{eq_integral}) can be evaluated by the saddle-point
method. Hence, we find\footnote{The exponent pre-factor
in~(\ref{eq_finalNumber}) is not required for computing the
asymptotic entropy, and therefore it is omitted.}
\BE\label{eq_finalNumber}
    \mathcal{N}(\beta,\gamma)\propto\lim_{K\rightarrow\infty}\exp{\big(Kg(a^{\ast},b^{\ast},\gamma,\beta)\big)},
\EE where $a^{\ast}$ and $b^{\ast}$ are found by the saddle-point
conditions, which yield the following equations \BEA
    \pd{g(a,b,\beta)}{a}&=&\beta^{-1}\big(\frac{(1-b)^{2}}{a}-1\big)+\gamma t\frac{Q'(t)}{Q(t)}=0\label{eq_saddle1},\\
    \pd{g(a,b,\beta)}{b}&=&\beta^{-1}\big(1-\frac{1-b}{a}\big)+\gamma\frac{1}{\sqrt{a\beta}}\frac{Q'(t)}{Q(t)}=0\label{eq_saddle2}.
\EEA The operator $Q'$ denotes a derivative of $Q$ w.r.t. its
argument. This set of saddle-point equations can be solved
numerically to obtain its fixed-points $a^{\ast}$, $b^{\ast}$ and
$t^{\ast}$.

Finally, substituting~(\ref{eq_finalNumber}) into
(\ref{eq_capacity}) the asymptotic entropy, in nats, is now easily
obtained \BEA\label{eq_C}
    H(\beta,\gamma)&=&g(a^{\ast},b^{\ast},\gamma,\beta)\nonumber\\&=&\frac{1}{\beta}\big(b^{\ast}-\frac{1}{2}+\frac{(1-b^{\ast})^{2}}{2a^{\ast}}+\frac{1}{2}\log{a^{\ast}}\big)+\gamma\log\big(2Q(t^{\ast})\big)+(1-\gamma)\log(2),
\EEA which, along with
equations~(\ref{eq_auxt}),(\ref{eq_saddle1})-(\ref{eq_saddle2}),
concludes our proof and forms the desired theorem.

\section{Results}
\begin{figure}[thb!]
\begin{center}
\psfig{file=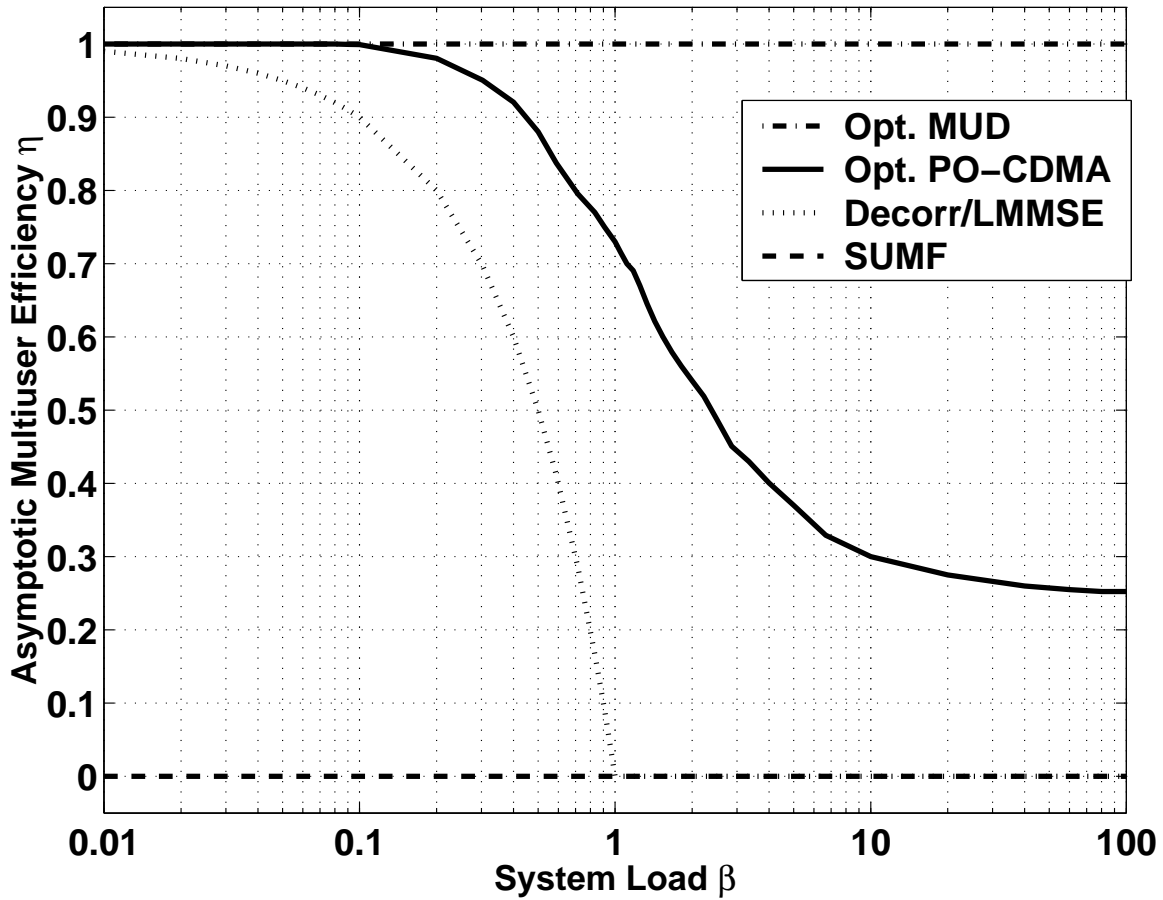,width=\textwidth}
\end{center}\vspace{-0.75cm}
\caption{Optimum AME of PO-CDMA compared to optimal MUD and linear
detectors.} \label{fig_eta}
\end{figure}

Fig.~\ref{fig_eta} presents the optimum AME of PO-CDMA
$\eta_{\textrm{opt}}(\beta)=\max_{\gamma}\eta(\beta,\gamma)$,
drawn via an exhaustive search over all possible values of the
active users fraction $\gamma$. Also drawn for comparison are the
AMEs of the single-user matched filter (SUMF), decorrelator, and
linear minimum mean square error (LMMSE)
detectors~\cite{BibDB:BookVerdu}.

Interestingly, $\eta_{\textrm{opt}}(\beta\lesssim0.1)$ is
practically equal to 1, which is the optimum AME of randomly
spread CDMA~\cite{BibDB:OAME}, obtained for the optimal MUD. For
large $\beta$ loads the optimum AME of the examined scheme
converges to 1/4, and not 0 as for the other sub-optimal
detectors. The corresponding optimal fraction $\gamma$ is found to
decrease from 1 to 1/2 as we increase $\beta$.

\vspace{6mm} \noindent {\large\bf Acknowledgment} \vspace{3mm}

\noindent This work was supported in part by the Advanced
Communications Center (ACC), Tel-Aviv University.

\end{document}